\documentclass[runningheads]{llncs}
\usepackage[show]{ed}
\usepackage{graphicx}
\usepackage{verbatim}
\usepackage{amsfonts}
\usepackage{amsmath}
\usepackage{stmaryrd}
\usepackage{listings}
\usepackage{enumitem}
\usepackage{upgreek}
\usepackage[english]{babel}
\usepackage{adjustbox}
\usepackage{xcolor}
\usepackage{wrapfig}
\usepackage[colorlinks=true]{hyperref}
\usepackage{breakurl}


\definecolor{ourblue}{rgb}{0.90,0.98,1.0}
\definecolor{ourpink}{rgb}{1.0,0.96,0.99}
\newcommand{\tog}{tog}

\newcommand{\bigdot}{\ensuremath{\boldsymbol{\cdot}}}
\newcommand{\sembr}[1]{\ensuremath{\llbracket  #1 \rrbracket}}

\begin{document}
\title{Leveraging the Information Contained in Theory Presentations}  
%
%
\author{Jacques Carette \and
William M. Farmer \and
Yasmine Sharoda}
\authorrunning{J. Carette \and W.M. Farmer \and Y. Sharoda}
%
\institute{Computing and Software, McMaster University, Hamilton, Canada 
\email{{\{carette,wmfarmer,sharodym\}}@mcmaster.ca}\\
\url{http://www.cas.mcmaster.ca/research/mathscheme/}}
\maketitle              
\begin{abstract}

A theorem prover without an extensive library is much less useful to its
potential users. Algebra, the study of algebraic structures, is a core
component of such libraries.  Algebraic theories also are themselves
structured, the study of which was started as Universal Algebra.  Various
constructions (homomorphism, term algebras, products, etc) and their properties
are both universal and constructive. Thus they are ripe for being automated.
Unfortunately, current practice still requires library builders to write
these by hand.
We first highlight specific redundancies in libraries of existing systems.
Then we describe a framework for generating these derived concepts from theory
definitions.  We demonstrate the usefulness of this framework on a test library
of $227$ theories.

\end{abstract}

\section{Introduction}
\label{sec:intro}
A theorem prover on its own is not nearly as useful for end-users as one
equipped with extensive libraries. Most users have tasks to perform
that are not related to new ideas in theorem proving. The larger the library
of standard material, the faster that users can just get
to work.  However building large libraries is currently very labor intensive.
Although some provers provide considerable automation for proof development, 
they do not the same for theory development.

This is the problem we 
continue~\cite{alhassy2019,carette2019big,CaretteOConnorTPC,carette2018building}
to tackle here, and that others~\cite{cohen2020hierarchy} have started
to look at as well. It is worthwhile noting that some programming languages
already provide interesting features in this direction.  For example,
Haskell~\cite{haskell} provides the \emph{deriving} mechanism that
lets one get instances for some classes ``for free''; recently, the
\emph{Deriving Via} mechanism~\cite{loeh2018derivingVia} has been 
introduced, that greatly amplifies these features. Some libraries, such as the 
one for \emph{Lens}~\cite{lensesLib}, use \emph{Template 
Haskell}~\cite{sheard2002TH} for the same purpose. 

Libraries of algebra define algebraic structures, constructions on these, and
properties satisfied by the structures and constructions.
While structures like \verb|Semigroup|, \verb|Monoid|,
\verb|AbelianGroup|, \verb|Ring| and \verb|Field| readily come to mind,
a look at compendiums~\cite{halleck,jipsen} reveals a much larger zoo
of hundreds of structures.

\begin{figure}
\footnotesize
\begin{tabular}{p{6.3cm} p{7cm}}
\begin{lstlisting}[mathescape]
$\text{\underline{Haskell}}$
class Semiring a => Monoid a 
 where 
  mempty :: a 
  mappend :: a -> a -> a 
  mappend = (<>) 
  mconcat :: [a] -> a 
  mconcat = 
   foldr mappend mempty 
$\text{\underline{Coq}}$
class Monoid {A : type}
 (dot : A $\to$ A $\to$ A)
 (one : A) : Prop := {
  dot_assoc : forall x y z : A, 
  (dot x (dot y z)) = 
  dot (dot x y) z
  unit_left : forall x, 
  dot one x = x 
  unit_right : forall x, 
  dot x one = x              
}
$\text{\textit{Alternative Definition:}}$
Record monoid := {
 dom : Type; 
 op : dom -> dom -> dom 
  where "x * y" := op x y; 
 id : dom where "1" := id; 
 assoc : forall x y z, 
  x * (y * z) = (x * y) * z; 
 left_neutral : forall x,   
  1 * x = x; 
 right_neutal : forall x,
  x * 1 = x; 
}
$\text{\underline{MathScheme}}$
Monoid := Theory { 
 U : type; 
 * : (U,U) $\to$ U; 
 e : U; 
 axiom right_identity_*_e : 
  forall x : U $\cdot$ (x * e) = x;
 axiom left_identity_*_e :  
  forall x : U $\cdot$ (e * x) = x;
 axiom associativity_* : 
  forall x,y,z : U $\cdot$ 
  (x * y) * z = x * (y * z); 
}
\end{lstlisting}
&
\begin{lstlisting}[mathescape]
$\text{\underline{Agda}}$
record Monoid c $\ell$ : 
   Set (suc (c $\sqcup$ $\ell$)) where 
 infixl 7 _$\bullet$_
 infix 4 _$\approx$_
 field 
  Carrier : Set c 
  _$\approx$_ : Rel Carrier $\ell$ 
  _$\bullet$_ : Op$_2$ Carrier 
  isMonoid : IsMonoid _$\approx$_ _$\bullet$_ $\varepsilon$ 
$\text{where }$ IsMonoid $\text{ is defined as }$
record IsMonid ($\bullet$ : Op$_2$) ($\varepsilon$ : A) 
 : Set (a $\sqcup$ $\ell$) where 
  field 
   isSemiring : IsSemiring $\bullet$ 
   identity : Identity $\varepsilon$ 
   identity$^l$ : LeftIdentity $\varepsilon$ $\bullet$ 
   identity$^l$ : proj$_1$ identity 
   identity$^r$ : Rightdentity $\varepsilon$ $\bullet$ 
   identity$^r$ : proj$_2$ identity           

$\text{\underline{MMT}}$
theory Semigroup : ?NatDed = 
 u : sort 
 comp : tm u $\to$ tm u $\to$ tm u 
  # 1 * 2 prec 40
 assoc : $\vdash$ $\forall$ [x, y, z]
  (x * y) * z = x * (y * z)    
 assocLeftToRight : 
  {x,y,z} $\vdash$ (x * y) * z 
          = x * (y * z) 
  = [x,y,z] 
   allE (allE (allE assoc x) y) z
 assocRightToLeft : 
  {x,y,z} $\vdash$  x * (y * z) 
           = (x * y) * z 
  = [x,y,z] sym assocLR 
theory Monoid : ?NatDed 
 includes ?Semigroup 
 unit : tm u # e 
 unit_axiom : $\vdash$ $\forall$ [x] = x * e = x       
\end{lstlisting}       
\end{tabular}  

\caption{Representation of \lstinline|Monoid| theory in different languages.}
\label{fig:monoid}
\end{figure}  
Picking~\verb|Monoid| as an example, it is
a structure with a carrier set, an associative binary operation and
an identity element for the binary operation. 
Different systems implement \verb|Monoid| in different ways
(see Figure~\ref{fig:monoid}). Other than layout and vocabulary, different
libraries also make more substantial choices:
\begin{itemize}
\item Whether declarations are arguments or fields.
\item The packaging structure --- whether theory, record, locale, etc.
\item The underlying notion of equality.
\end{itemize}

Some of these choices are mathematically irrelevant --- in the sense that
the resulting theories can be proved to be equivalent, internally or
externally --- while others are more subtle, such as the choice of
equality.

A useful construction on top of \verb|Monoid| is the homomorphism between two
of its instances, which maps elements of the carrier of the first instance to
that of the second one such that structure is preserved. For an operation
\lstinline|op| and a function \lstinline|hom|, the preservation axiom has the
form
\begin{lstlisting}[mathescape]
hom (op x$_1$ .. x$_n$) = op (hom x$_1$) .. (hom x$_n$)
\end{lstlisting} 
One can see that this definition can be ``derived'' from that of \verb|Monoid|.
And that, in fact, this derivation is uniform in the ``shape'' of the
definition of \verb|Monoid|, so that this construction applies to any
single-sorted equational theory. This observation is one of the cornerstones of
Universal Algebra~\cite{whitehead1898treatise}.

There are other classical constructions that can also be generated. This poses
a number of questions:
\begin{itemize}
\item What other information can be generated from theory presentations? 
\item How would this affect the activity of library building? 
\item What pieces of information are needed for the system to generate
particular constructions? 
\end{itemize}
Theories written in equational logic that describe algebraic
structures are rich in implicit information that can be extracted automatically.

\begin{sloppypar}
There are obstacles to this automation.  For example, definitional and
``bundling'' choices can make reuse of definitions from one project in
another with different aims difficult.  Thus users resort to redefining
constructs that have already been formalized. We then end up with multiple
libraries for the same topic in the same system.  For example, there are at
least four algebra libraries in Coq
\cite{Gonthier2009,Geuvers2002,coq-contribs-algebra,Spitters2010},
and even more for Category Theory~\cite{spivak2014coqcats}. In 
\cite{Gonthier2009}, the authors
mention, referring to other libraries:  
\begin{quote}
  ``In spite of this body of prior work, however, we have found it
difficult to make practical use of the algebraic hierarchy in our project to
formalize the Feit-Thompson Theorem in the Coq system."
\end{quote}
\end{sloppypar}

Universal
Algebra~\cite{meinke1991universal,sankappanavar1981course,whitehead1898treatise}
provides us with tools and abstractions well-suited to this task.  It is
already used in providing semantics and specifications of computer systems
\cite{denecke2002universal,ehrig2012fundamentals,sannella2012foundations} and
has been formalized in Coq~\cite{capretta99} and Agda~\cite{Gunther2018Agda}. 
We use Universal Algebra abstractions as basis for our framework to automate
the  generation of useful information from the definition of a 
theory. We use Tog to realize our framework\footnote{The implementation is
available at \url{https://github.com/ysharoda/tog}.}. Tog is a small
implementation of a dependent type theory, in the style of Agda, Idris and Coq.
It serves well as an abstraction over the design details of different systems.
Studying theory presentations at this level of abstraction is the first step to
generating useful constructions for widely used systems, like Agda, Coq,
Isabelle and others. 
  
In Section~\ref{sec:redundancies} we highlight some of the redundancies in
current libraries.  We present our framework for mechanizing the generation of
this information in Section~\ref{sec:free}.  We follow this with a discussion of
related work in Section~\ref{sec:related_work} and a conclusion and future work
in Section~\ref{sec:conc}.

\section{Algebra in Current Libraries}
\label{sec:redundancies}

Our first observation is that current formalizations of Algebra contain quite a
bit of information that is ``free'' in the sense that it can be
mechanically generated from basic definitions. For example, given a theory
X, it is mechanical to define X-homomorphisms. To do this within a system
is extremely difficult, as it would require introspection and for theory
\emph{definitions} to be first-class citizens, which is not the case for any
system based on type-theory that we are aware of. Untyped systems
in the Lisp tradition do this routinely, as does Maude~\cite{Maude}, which
is based on \emph{rewriting logic}; the downside is that there is no
difference between meaningful and meaningless transformations in these
systems, only between ``runs successfully'' and ``crashes''. However,
these constructions are fully typeable and, moreover, are not system-specific
(as they can be phrased meta-theoretically within Universal
Algebra), even though an implementation has to be aware of the syntactic
details of each system.

Lest the reader think that our quest is a little quixotic, we first look at
current libraries from a variety of systems, to find concrete examples of
human-written code that could have been generated. We look at Agda,
Isabelle/HOL and Lean in particular. More specifically, we look at
\href{https://github.com/agda/agda-stdlib/releases/tag/v1.3}
{version 1.3 of the Agda standard library},
the \href
{https://isabelle.in.tum.de/website-Isabelle2019/dist/library/HOL/HOL-Algebra/index.html} 
{2019 release of the Isabelle/HOL library}
and 
\href
{https://github.com/leanprover-community/mathlib/releases/tag/snapshot-2019-10}
{Lean's mathlib}, where we link to the proper release tag.

We use the theory \verb|Monoid| as our running example, and we 
highlight the reusable components that the systems use to make writing the
definitions easier and more robust. 

\subsection{Homomorphism}
How do the libraries of our three systems\footnote{We do not have enough room
to give an introduction to each system; hopefully each system's syntax is
clear enough for the main ideas to come through.}
represent homomorphism?

\subsubsection{Agda}
defines \verb|Monoid| homomorphism, indirectly, in two ways. First,
a predicate encapsulating the proof obligations is defined, which is
layered on top of the predicate for 
\href
{https://github.com/agda/agda-stdlib/blob/4a8d8f5ffbdbd967ca1bb708895ea63709e0063d/src/Algebra/Morphism.agda} 
{Semigroup homomorphism}.
This is then used to define homomorphisms themselves.

\begin{lstlisting}[mathescape]
module _ {c$_1$ $\ell_1$ c$_2$ $\ell_2$}
         (From : Monoid c$_1$ $\ell_1$)
         (To   : Monoid c$_2$ $\ell_2$) where

 private
  module F = Monoid From
  module T = Monoid To

 record IsSemigroupMorphism ($\llbracket$_$\rrbracket$:Morphism)
        : Set(c$_1$ $\sqcup$ $\ell_1$ $\sqcup$ c$_2$ $\sqcup$ $\ell_2$) where 
  field
   $\llbracket\rrbracket$-cong : $\llbracket$_$\rrbracket$ Preserves F._$\approx$_ $\to$ T._$\approx$_
   $\bigdot$-homo  : Homomorphic$_2$ $\llbracket$_$\rrbracket$ F._$\bigdot$_ T._$\bigdot$_
   $\cdots$
 record IsMonoidMorphism ($\llbracket$_$\rrbracket$:Morphism)
        : Set(c$_1$ $\sqcup$ $\ell_1$ $\sqcup$ c$_2$ $\sqcup$ $\ell_2$) where 
  field
   sm-homo : IsSemigroupMorphism F.semigroup T.semigroup $\llbracket$_$\rrbracket$
   $\varepsilon$-homo   : Homomorphic$_0$ $\llbracket$_$\rrbracket$ F.$\varepsilon$ T.$\varepsilon$

 open IsSemigroupMorphism sm-homo public
\end{lstlisting}
There are many design decisions embedded in the above definitions. These
decisions are not canonical, so we need to understand them to later
be able to both abstract them out and make them variation points in our
generator. Namely, these decisions are:
\begin{itemize}
  \item The choice of which declarations are parameters and which are fields.
  The monoids (\verb|From| and \verb|To|) over which we define homomorphism
  are parameters, not fields, as is the function
  \lstinline[mathescape]|$\llbracket$_$\rrbracket$|.
  \item The preservation axioms can be defined based on their arity
   patterns, as type-level function such as
\lstinline[mathescape]|Homomorphic$_{2}$|:
  \begin{lstlisting}[mathescape]
  Homomorphic$_{2}$ : (A $\to$ B) $\to$ Op$_2$ A $\to$ Op$_2$ B $\to$ Set _ 
  Homomorphic$_{2}$ $\llbracket$_$\rrbracket$ _$\bigdot$_ _$\circ$_ =
    $\forall$ x y $\to$ $\sembr{\text{ x } $\bigdot$ \text{ y }}$ $\approx$ ($\sembr{\text{ x }}$ $\circ$ $\sembr{\text{ y }}$)
  \end{lstlisting}
  The library also provides shortcuts for $0$-ary and $1$-ary function
  symbols, the most common cases.
  \item The definition of structures over setoids. Thus equalities need 
  to be preserved, and that is what the 
  \lstinline[mathescape]|$\llbracket\rrbracket$-cong| axiom states.
\end{itemize}

\subsubsection{Isabelle/HOL}
provides the following definition of \href{https://isabelle.in.tum.de/website-Isabelle2019/dist/library/HOL/HOL-Algebra/Group.html}{monoid homomorphism}:
\begin{lstlisting}[mathescape]
definition
hom :: "_ $\Rightarrow$ _ $\Rightarrow$ ('a $\Rightarrow$ 'b) set" where
  "hom G H =
  {h $\cdot$ h $\in$ carrier G $\to$ carrier H $\wedge$ 
  ($\forall$ x $\in$ carrier G $\cdot$ $\forall$ y $\in$ carrier G $\cdot$ 
    h (x $\oplus_\text{G}$ y) = h x $\oplus_\text{H}$ h y)}"
\end{lstlisting}
The reader might notice a discrepancy in the above: unit preservation is
missing.  The Isabelle library does not provide this version.
There is, however, a proof that such a multiplication-preserving homomorphism
necessarily maps the source unit to a unit of the image (sub)monoid, but that
unit is not necessarily that of the full image. The above definition is also
used to define group homomorphism and other structures. We consider
this to be missing information in the library. 

\subsubsection{Lean}\hspace{-0.4em}'s
definition of \href{https://github.com/leanprover-community/mathlib/blob/3c58f160fd51ebf989138ed7c8981f821f08f860/src/algebra/group/hom.lean}
{monoid homomorphism}
is the one that most resembles the one found in textbooks. 
\begin{lstlisting}[mathescape]
structure monoid_hom (M : Type*) (N : Type*) 
 [monoid M] [monoid N] :=
  (to_fun : M $\to$ N)
  (map_one' : to_fun 1 = 1)
  (map_mul' : $\forall$ x y, to_fun (x * y) = to_fun x * to_fun y)
\end{lstlisting}
However, in the same file, there is another definition of \verb|add_monoid_hom| that looks ``the same'' up to renaming. This points
to a weakness of Lean: there is no renaming operation on 
\verb|structure|, and for a \verb|Ring| to contain two ``monoids'', one
is forced to duplicate definitions. This redundancy is unpleasant.

\subsection{Term Language}

The ``term language'' of a theory is the (inductive) data type
that represents the syntax of well-formed terms of that theory,
along with an interpretation function from \emph{expressions} 
to the carrier of the (implicitly single-sorted) given theory, i.e.
its denotational semantics.

In Agda, the definition of \lstinline|Monoid| term language is straightforward:
\begin{lstlisting}[mathescape]
data Expr (n : $\mathbb{N}$) where 
  var : Fin n $\to$ Expr n 
  id : Expr n 
  _$\oplus$_ : Expr n $\to$ Expr n $\to$ Expr n 
\end{lstlisting}

Defining the interpretation function requires the concept of an environment.
An environment associates a value to every variable, and the semantics
associates a value (of type \verb|Carrier|) to each expression of \verb|Expr|.
\begin{lstlisting}[mathescape]
Env : Set _ 
Env = $\lambda$ n $\rightarrow$ Vec Carrier n 

$\llbracket$_$\rrbracket$ : $\forall$ {n} $\to$ Expr n $\to$ Env n $\to$ Carrier 
$\llbracket$ var x $\rrbracket$ $\upvarrho$ = lookup $\upvarrho$ x 
$\llbracket$ id $\rrbracket$ $\upvarrho$ = $\epsilon$ 
$\llbracket$ e$_1$ $\oplus$ e$_2$ $\rrbracket$ $\upvarrho$ = $\llbracket$ e$_1$ $\rrbracket$ $\upvarrho$ $\cdot$ $\llbracket$ e$_2$ $\rrbracket$ $\upvarrho$ 
\end{lstlisting}

In Agda, these definitions are not found with the definitions of the
algebraic structures themselves, but rather as part of the
\emph{Solver} for equations over that theory. Here, we find more
duplication, as the above definitions
are repeated for the following three highly related structures: 
\href{https://github.com/agda/agda-stdlib/blob/4a8d8f5ffbdbd967ca1bb708895ea63709e0063d/src/Algebra/Solver/Monoid.agda}
{\lstinline|Monoid|},
\href{https://github.com/agda/agda-stdlib/blob/4a8d8f5ffbdbd967ca1bb708895ea63709e0063d/src/Algebra/Solver/CommutativeMonoid.agda}
{\lstinline|CommutativeMonoid|}
and 
\href{https://github.com/agda/agda-stdlib/blob/4a8d8f5ffbdbd967ca1bb708895ea63709e0063d/src/Algebra/Solver/IdempotentCommutativeMonoid.agda}
{\lstinline|IdempotentCommutativeMonoid|}.

Despite its usefulness, we were not able to find the definition of the term
language of a theory in Isabelle/HOL or Lean.  

\subsection{Product}
Until recently, there was no definition of the product of algebraic
structures in the Agda library.  A 
\href{https://github.com/agda/agda-stdlib/pull/1109}{recent pull request}
has suggested adding these, along with other constructions.  The
following hand-written definition has now been added:
\begin{lstlisting}[mathescape]
rawMonoid : RawMonoid c c$\ell$ $\to$ RawMonoid d d$\ell$ $\to$ 
RawMonoid (c $\sqcup$ d) (c$\ell$ $\sqcup$ d$\ell$)
rawMonoid M N = record
  { Carrier = M.Carrier $\times$ N.Carrier
  ; _$\approx$_ = Pointwise M._$\approx$_ N._$\approx$_
  ; _$\bigdot$_ = zip M._$\bigdot$_ N._$\bigdot$_
  ; $\varepsilon$ = M.$\varepsilon$ , N.$\varepsilon$
  }
  where
  module M = RawMonoid M
  module N = RawMonoid N
\end{lstlisting}
These could have been mechanically generated from the definition
of \verb|Monoid|.

Both 
\href{https://isabelle.in.tum.de/website-Isabelle2019/dist/library/HOL/HOL-Algebra/Group.html}
{Isabelle/HOL}
and 
\href{https://github.com/leanprover-community/mathlib/blob/3c58f160fd51ebf989138ed7c8981f821f08f860/src/algebra/pi_instances.lean}
{Lean}
provide definitions of product algebras for monoids, which we omit for space.
It is worth mentioning that the Lean library has $15$ definitions for products
of structures that look very similar and could be generated. 

\subsection{More Monoid-Based Examples}

We have presented three concrete examples, based on monoid, of
human-written code in current libraries that could have instead been
generated. There are many more that could be, although these are sparsely found
in current libraries. We continue to use monoid as our guiding example, and
also briefly discuss how they can be generalized to a larger algebraic context
and why they are useful. These are presented in a syntax that closely
resembles that of Agda (and is formally Tog syntax), which should be
understandable to anyone familiar with dependently-typed languages.

\subsubsection{Trivial Submonoid.}
Given a monoid \lstinline|M|, we can construct the trivial monoid,
also called the zero monoid\footnote{as it is both initial and terminal
in the corresponding Category} (containing only the identity element) in the
same language as \lstinline|M|.

\begin{lstlisting}[mathescape]
record TrivialSubmonoid {A : Set} (M : Monoid A) : Set 
where
  constructor trivialSubmonoid
  field
    singleton : {x : A} $\to$ x == M.e
\end{lstlisting}
One can easily proceed to show that this predicate on a monoid
induces a new (sub)monoid.  In fact, we do not need associativity
for this; in other words, already a unital magma induces a trivial monoid.

\subsubsection{Flipped Monoid.} 
Given a monoid \lstinline|M|, we can construct a new monoid where the
binary operation is that of \lstinline|M| but applied in reverse order.

The construction here is direct, in that the result is a \lstinline|Monoid|.
\begin{lstlisting}[mathescape]
record FlippedMonoid : {A : Set} $\to$ Monoid A $\to$ Monoid A
record FlippedMonoid m = {
  A = M.A,
  e = M.e,
  op = (x y : A) $\to$ M.op y x,
  lunit = M.runit,
  runit = M.lunit,
  assoc = sym M.assoc
}
\end{lstlisting}
This example can be generalized from a monoid to a magma.

\subsubsection{Monoid Action.}
This example constructs, from a Monoid \lstinline|M| and a set 
\lstinline|B|, a monoid action of \lstinline|M| on \lstinline|B|.

\begin{lstlisting}[mathescape]
record MonoidAction {A : Set} (M : Monoid A) 
                    (B : Set) : Set where
  constructor monoidAction
  field
    act     : A $\to$ B $\to$ B
    actunit : {b : B} $\to$ (act M.e b) == b
    actop   : {x y : A} $\to$ {b : B} $\to$ 
                (act (M.op x y) b) == (act x (act y b))
\end{lstlisting}
Monoid actions are extremely useful for expressing ideas in group
theory, and in automata theory. They are only defined in the presence of a
monoid structure, which can be easily checked at the meta level. 

\subsubsection{Subsets Action.}

The fourth example construct, from a Monoid \lstinline|M|, the 
monoid on the subsets of \lstinline|M|.  Note that the following
is pseudo-code written in an imagined Set-theoretic extension
of dependent type theory.

\begin{lstlisting}[mathescape]
record SubsetsAction {A : Set} (M : Monoid A) : Set 
where
  constructor subsetsAction
  field
    S      : (powerset A)
    e'     : S
    op'    : S $\to$ S $\to$ S
    e'def  : e' == {M.e} 
    op'def : {x y : S} $\to$ (op' x y) 
               == {(M.op a b) | a $\in$ x and b $\in$ y}
\end{lstlisting}
The subsets monoid is used extensively in automata theory and group
theory.

The above can also be written as a construction of a new monoid,
in dependent type theory, where the carrier is the set of unary
relations on $A$.

\subsubsection{Monoid Cosets.}
The next example constructs, from a Monoid $M$, the cosets of $M$.
This is also pseudo-code, as above.

\begin{lstlisting}[mathescape]
record MonoidCosets {A : Set} (M : Monoid A) : Set 
where
  constructor monoidCosets
  field
    S      : (powerset A)
    e'     : S
    op'    : A $\to$ S $\to$ S
    e'def  : e' == {M.e}
    op'def : {a : A} $\to$ {x : S} $\to$ (op' a x) 
               == {(M.op a b) | b $\in$ x}
\end{lstlisting}
Monoid cosets are extensively used in group theory.

\section{Constructions for Free!}
\label{sec:free}
A meta-theory (either a logic or a type theory) provides us with a concrete
language in which to represent axiomatic theories. Through having a uniform
syntactic representation of the components of axiomatic theories, we can
manipulate them, and eventually generate new ones from them.

Our meta-theory is Martin-L\"{o}f Type Theory, as implemented in
Tog~\cite{tog}. Tog is developed by the implementors of Agda for the purpose of
experimenting with new ideas in (implementations of) dependent type
theories. It has mainly been used to experiment with type checking through unification~\cite{mazzoli2016type}. Tog is minimalistic, and serves our purpose of being
independent of the design details of many of the large proof languages.
It also gives us a type checker.  

The following implementation details of Tog are worth pointing out:
\begin{itemize}
\item It has one universe \lstinline|Set|, which is the kind of all
sorts.
\item Functions are represented as curried lambda expressions:
\lstinline|Fun Expr Expr|.
\item Axioms are represented as \lstinline[mathescape]|$\Uppi$|-types: 
  \lstinline|Pi Telescope Expr|. They use the underlying propositional 
  equality: \lstinline|Eq Expr Expr|.  
\item Theories are represented as parameterized dependent records,
  $\Upsigma$-types. 
 \begin{itemize}
   \item A parameter to the record has the type \lstinline|Binding|. It can be
  hidden using \lstinline|HBind [Arg] Expr|, or explicit using
  \lstinline|Bind [Arg] Expr|. 
   \item A declaration within the record has the type 
         \lstinline|Constr Name Expr|.
 \end{itemize}
\end{itemize} 

In Universal Algebra, an algebraic theory consists of sorts, function symbols
(with their arities) and a list of axioms, often denoted as a theory
\lstinline|T| having three components \lstinline|(S,F,E)|.  We assume a single
sort.  This can be internalized, in the Haskell implementation of Tog, as
\begin{lstlisting}
data EqTheory = EqTheory  {
  thryName    :: Name_   ,
  sort        :: Constr  , 
  funcTypes   :: [Constr],
  axioms      :: [Constr],
  waist       :: Int     }
\end{lstlisting}
where: 
\begin{itemize}
  \item \lstinline|sort|, \lstinline|funcTypes|, and \lstinline|axioms| are
treated as elements of a telescope~\cite{de1991telescopic}. Therefore, the
order in which they are defined matters. 
  \item The \lstinline|waist| is a number referring to how many of the
declarations within the telescope are parameters. The notation is taken
from~\cite{alhassy2019}. This information is needed in generating some
constructions, like homomorphism. 
\end{itemize}

Given a Tog record type that exhibits an equational theory structure, like that
of \lstinline|Monoid| in Section~\ref{sec:intro}, we convert it into an
instance of \verb|EqTheory|. We, then, proceed with generating useful
information from the theory. Finally, we convert this information into Tog
records and data types, so they can be type checked by Tog, i.e. our approach
builds on Tog, without changing its syntax or type checker. In the sequel of
this section, we describe the constructions we generate. 

\subsection{Signature}
\label{sec:free_sig}
Given a theory \lstinline|T = (S,F,E)|, the signature of the theory is
\lstinline|Sig(T) = (S,F)|. A signature is obtained from an \verb|EqTheory| as follows: \begin{lstlisting}
signature_ :: Eq.EqTheory -> Eq.EqTheory
signature_ = 
  over Eq.thyName (++ "Sig") . set Eq.axioms [] .  gmap ren
\end{lstlisting}
For a theory with name \lstinline|X|, the signature is an \verb|EqTheory| with the name \lstinline|XSig| and an empty axioms list. The theory and its signature exists in the same module. Tog requires that they have different field names. We use \lstinline|gmap ren| to apply this renaming. We discuss this in more details in Section~\ref{subsec:free_disc}. 

\subsection{Product Algebra}
\label{sec:free_prod}
Given a theory~\lstinline|T = (S,F,E)|, we obtain the product theory 
\lstinline[mathescape]|Prod(T) = (S$\times$S, F$^\prime$, E$^\prime$)| by
replacing each occurrence of the type
\lstinline|S| by \lstinline[mathescape]|S$\times$S|. 
The modification to the function symbols and axioms is straightforward. 
\begin{lstlisting}
productThry :: Eq.EqTheory -> Eq.EqTheory
productThry t =   
  over Eq.thyName (++ "Prod") $
  over Eq.funcTypes (map mkProd) $
  over Eq.axioms (map mkProd) $ 
  gmap ren t
\end{lstlisting}
Similar to what we did with signatures, the \lstinline|ren| function renames the fields of the input theory. \lstinline|mkProd| changes the sort to be an instance of \lstinline|Prod|, with the sort of the input theory as the type parameter. 

\subsection{Term Language}
\label{sec:free_lang}
For a theory \lstinline|T = (S,F,E)|, the
closed term language is generated by converting every function symbol to a
constructor, with the same arity. The axioms are dropped.
\begin{lstlisting}
termLang t =
 let constructors = 
  gmap (ren (getConstrName $ t^.Eq.sort) nm) $ t^.Eq.funcTypes
 in Data (mkName $ t^.thyName ++ "Lang") NoParams $
  DataDeclDef setType constructors
\end{lstlisting}
Constructors are generated by substituting the name of the language type for a
sort \lstinline|A|. Term languages are realized as Tog data declarations using
the constructor \lstinline|Data|.

Generating the closed term language is a first step to generating an open term
language (i.e. a term language parametrized by a type of variables), 
and an interpreter. 

For some kinds of axioms, namely those that can be \emph{oriented}, we
can turn these into \emph{simplification rules}, i.e. into (unconditional)
rewrite rules. The resulting simplifier can be shown to be meaning preserving.
These two pieces, the evaluator and simplifier, can be attached to each other
to form a \emph{partial evaluator}, using the ``finally
tagless''\cite{carette2009finally} method.  Eventually, we would like to be
able to automate the majority of the hand-written code for a generative
geometry library \cite{carette2011generative}, which is indeed quite amenable
to such techniques. Unfortunately, the details will have to wait for a
future paper.


\subsection{Homomorphism}
\label{sec:free_hom}
For a theory \lstinline|T = (S,F,E)|,
with instances \lstinline[mathescape]|T$_1$| and
\lstinline[mathescape]|T$_2$|, the homomorphism
of \lstinline|T| consists of 
\begin{enumerate}
\item a function mapping the carrier of 
\lstinline[mathescape]|T$_1$| to that of
\lstinline[mathescape]|T$_2$|,
\item a set of axioms asserting that operations (i.e. elements of
\lstinline|F|) are preserved.
\end{enumerate}
Our definition of homomorphism is parameterized by the instances
\lstinline[mathescape]|T$_1$| and \lstinline[mathescape]|T$_2$|. The parameters
of \lstinline|T|, if \lstinline|waist| $>0$, are lifted out as parameters to
the resulting homomorphism, and used to define the instances of the theory.  
\begin{lstlisting}
homomorphism :: Eq.EqTheory -> Decl
homomorphism t =
 let nm = t ^. Eq.thyName ++ "Hom"
     a = Eq.args t 
     (psort,pfuncs,_) = mkPConstrs t
     ((i1, n1), (i2, n2)) = createThryInsts t
     homFnc = genHomFunc psort n1 n2
     axioms = map (oneAxiom fnc psort n1 n2) pfuncs 
 in Record (mkName nm)
    (mkParams $ (map (recordParams Bind) a) ++ [i1,i2])
    (RecordDeclDef setType 
                 (mkName  $ nm ++ "C") 
                 (mkField $ fnc : axioms))
\end{lstlisting}
The \verb|genHomFunc| function generates the homomorphism function. Each
preservation axiom is created using the \lstinline|oneAxiom| function. 

Other kinds of morphisms can also be generated by providing more axioms to
describe properties of the functions. For example a monomorphism would have the
same definition with one more axiom stating that the function is injective. 
An endomorphism is a self-homomorphism, and thus can be parametrized by a
single theory.

\subsection{Discussion}
\label{subsec:free_disc}
The above are a small sample of what can be done. We've found at least
$30$ constructions that should be amenable to such a treatment and are
currently implementing them, including quotient algebras and induction axioms.  
Figure~\ref{fig:generated} shows the generated constructions. The input is the
theory of \verb|Monoid| represented as a Tog record type (illustrated on the
left with the blue background). For this, we generate the four constructions
discussed above (illustrated with pink background). 
\begin{figure}[htb]
    \begin{adjustbox}{width=\columnwidth,center}
    \begin{tabular}{p{7cm} p{2mm} p{7cm}}        
\lstset{backgroundcolor=\color{ourblue}}
\begin{lstlisting}[mathescape]
record Monoid (A : Set) : Set 
 where
  constructor monoid
  field
   e  : A
   op : A $\to$ A $\to$ A
   lunit : {x : A} $\to$ (op e x) == x
   runit : {x : A} $\to$ (op x e) == x
   assoc : {x y z : A} $\to$ 
     op x (op y z) == op (op x y) z
\end{lstlisting}
\lstset{backgroundcolor=\color{ourpink}}
\begin{lstlisting}[mathescape]
record MonoidHom 
    (A1 : Set) (A2 : Set)
    (Mo1 : Monoid A1)
    (Mo2 : Monoid A2) : Set where
  constructor MonoidHomC
  field
    hom : A1 $\to$ A2
    pres-e : hom (e Mo1) == e Mo2
    pres-op : 
      (x1 : A1) (x2 : A1) $\to$
      hom (op Mo1 x1 x2) 
      == op Mo2 (hom x1) (hom x2)
\end{lstlisting}
&
&
\lstset{backgroundcolor=\color{ourpink}}
\begin{lstlisting}[mathescape]
data MonoidLang : Set where
 eL : MonoidLang
 opL : MonoidLang $\to$ MonoidLang 
      $\to$ MonoidLang


record MonoidSig (AS : Set) : Set 
 where
  constructor MonoidSigSigC
  field
   eS : AS
   opS : AS $\to$ AS $\to$ AS    
    
    
record MonoidProd (AP : Set) 
   : Set 
 where
  constructor MonoidProdC
  field
   eP : Prod AP AP
   opP : Prod AP AP $\to$ Prod AP AP 
       $\to$ Prod AP AP
   lunit_eP : (xP : Prod AP AP) 
       $\to$ opP eP xP == xP
   runit_eP : (xP : Prod AP AP) 
       $\to$ opP xP eP == xP
   associative_opP : 
       (xP : Prod AP AP) 
       (yP : Prod AP AP) 
       (zP : Prod AP AP) 
       $\to$ opP (opP xP yP) zP 
         == opP xP (opP yP zP)   
\end{lstlisting}
\end{tabular}  
\end{adjustbox}

    \caption{The generated constructions from \lstinline|Monoid| theory}
    \label{fig:generated}
\end{figure}
The names of carriers \lstinline[mathescape]|A$_1$| and
\lstinline[mathescape]|A$_2$|, names of instances
\lstinline[mathescape]|Mo$_1$| and  \lstinline[mathescape]|Mo$_2$| are machine
generated based on the names used by the input theory, which are given by the
user. A somehow unpleasant restriction is that all field names need to be
distinct, even if the fields belong to different records. That is the reason we
have names like \lstinline|eL| in \lstinline|MonoidLang| and \lstinline|eS| in
\verb|MonoidSig|. This is still a minor inconvenience, given that we are
working on an abstract level, from which more readable and usable code will be
generated. 

\section{Related Work}
\label{sec:related_work}
Many algebraic hierarchies have been developed before.  \cite{Geuvers2002}
documents the development of the algebra needed for proving the fundamental
theorem of algebra. \cite{Gonthier2009} formalizes the same
knowledge in Coq, but suggests a packaging structure alternative to telescopes,
to support multiple inheritance. \cite{cohen2020hierarchy} addresses the
important problem of library maintainability, especially when dealing with
changes to the hierarchy. We have proposed an alternate solution in
\cite{carette2018building}, based on the categorical structures already
present in dependent type theories.

The algebraic library of Lean~\cite{lean2019} is of particular interest,
as its developers are quite concerned with automation. But this
automation, also done via meta-programming, is largely oriented to proof
automation via tactics.  We instead focus on
automating the generation of structures. 

Universal Algebra constructions are grounded in set theory, yet is
nevertheless quite constructive. It has been formalized in
Coq~\cite{capretta99,Spitters2010} and Agda~\cite{Gunther2018Agda}.
\cite{Spitters2010} is notable for the use of type classes to
formalize the algebraic hierarchy. 

While the work in interactive provers has been mainly manual, the
programming languages community has been actively investigating the
generation of various utilities derived from the definition of algebraic
data types.  Haskell's \emph{deriving} mechanism has already been mentioned.
This has been greatly extended twice, first
in~\cite{loeh2010genericDeriving}, to allow more generic deriving, and then
in~\cite{loeh2018derivingVia} allowing the users to define new patterns. 
The usefulness of these mechanisms has been of great inspiration to us.
We would like to provide similar tools for library developers of
interactive proof systems.

\section{Conclusion and Future Work}
\label{sec:conc}
Building large libraries of mathematical knowledge can greatly enhance
the usefulness of interactive proof systems.
Currently, the larger the library, the more labor intensive it
becomes. We suggest automating some of the definitions of concepts
derivable via known techniques.  We have tested our implementation on
a library of $227$ theories, including~\lstinline|Ring| and~\lstinline|BoundedDistributedLattice|,
built using the tiny theories
approach~\cite{mathscheme2011experiments} and the combinators
of~\cite{carette2018building}. A theory defined declaratively using the combinators elaborate into a Tog record, which is then manipulated to generate the constructions presented in Section~\ref{sec:free}. From the declarative description of the $227$
theories, we were able to generate a much larger library which contains $1132$ definitions and, when
pretty-printed, spanned $14811$ lines, containing theories and data types
representing the structures we discussed in Section~\ref{sec:free}.
We are adding more derived theories, and can then get a multiplicative
factor, as each time we do, we get $227$ new theories.

While the knowledge representable in single-sorted equational logic is
still impressive (e.g. it covers most of Algebra), we are also
interested in generating the same structures (and more) for
theories represented in more sophisticated logics~\cite{meinke1992higherTypes},
such as category theory represented in dependent type theory.  

We currently generate all constructions for all theories in a given library. As
more structures get generated, we would want to give developers
more control over what to generate. Thus we intend to provide
a scripting language for referring to theories, or groups of theories, and
specifying what constructions to apply. This could also include an
``on demand'' version, similar to how the deriving
mechanism of Haskell works. We are also interested in generating morphisms, as
explained in~\cite{LittleTheories}, between theories. Even for our
constructions, some of these morphism are not obvious, but are needed to
transport results.

We envision using our current implementation as a meta-language to
generate definitions for existing, full-featured systems, 
such as Isabelle/HOL and Agda. To achieve this, we will need to
reintroduce certain details (such as notations) that we elided. The
scripting language described above will need to be extended to cover
different kinds of \emph{design decisions}.

We envision a framework in which the contents of the
library can be defined succinctly, and elaborated to a
large reusable and flexible body of standardized mathematics
knowledge.
\bibliographystyle{plain}
\bibliography{paper.bib}
\end{document}